# Do altmetrics correlate with the quality of papers?

# A large-scale empirical study based on F1000Prime data

Lutz Bornmann* and Robin Haunschild **

* bornmann@gv.mpg.de

Division for Science and Innovation Studies, Administrative Headquarters of the Max Planck Society, Hofgartenstr. 8, 80539 Munich, (Germany)

** R.Haunschild@fkf.mpg.de

Max Planck Institute for Solid State Research, Heisenbergstr. 1, 70569 Stuttgart, (Germany)

**Abstract**

In this study, we address the question whether (and to what extent, respectively) altmetrics are related to the scientific quality of papers (as measured by peer assessments). Only a few studies have previously investigated the relationship between altmetrics and assessments by peers. In the first step, we analyse the underlying dimensions of measurement for traditional metrics (citation counts) and altmetrics – by using principal component analysis (PCA) and factor analysis (FA). In the second step, we test the relationship between the dimensions and quality of papers (as measured by the post-publication peer-review system of F1000Prime assessments) – using regression analysis. The results of the PCA and FA show that altmetrics operate along different dimensions, whereas Mendeley counts are related to citation counts, and tweets form a separate dimension. The results of the regression analysis indicate that citation-based metrics and readership counts are significantly more related to quality, than tweets. This result on the one hand questions the use of Twitter counts for research evaluation purposes and on the other hand indicates potential use of Mendeley reader counts.

# 1 Introduction

Altmetrics denote non-traditional metrics which represent an alternative form of impact measurement instead of using citations [1]. Altmetrics usually not only cover activities on social media platforms including, for example, mentions in blog posts, readership counts on Mendeley, shares on Facebook, and tweets, but also mentions in mainstream media and policy documents [2,3]. Thus, altmetrics refer to a heterogeneous set of metrics which is gaining increasing popularity amongst researchers, research communicators, publishers, and research funders [4]. For example, Wiley, Springer, BioMed Central and the Nature Publishing Group are adding altmetrics to papers in their collections. An overview of proposed definitions of altmetrics in the literature can be found in Erdt et al. [5]; the history of altmetrics is described by Wilsdon et al. [1]. One important reason for the use of altmetrics in research evaluation is the measurement of wider research impact: research with important societal or cultural impacts may be undervalued if assessed with citation-based indicators. Thus, other data sources are needed "if quantitative indicators are to be used to aid the evaluation of the wider impacts of academic research" [6].

Although altmetrics are already in use for research evaluation purposes [7], the main problem is that "it is still not clear what general conclusions can be drawn when an article is frequently mentioned within the social web" [8]. Zahedi et al. [9] ask the following (unanswered) questions in the context of altmetrics use in research evaluation: "What does it reflect when an item is saved/added by several users to their libraries? Also, what does it mean that an item is mentioned in Wikipedia, CiteULike, Twitter and any other social media platform?".

Following earlier studies by Bornmann [10] and Bornmann [11], we address therefore in this study the question whether (and to what extent, respectively) altmetrics are related to the scientific quality of papers (as measured by peer assessments). Some years ago,

F1000Prime was launched as a new type of post-publication peer-review system, in which around 5,000 experts ("Faculty members") were asked to identify, assess, and comment on interesting papers they read [12]. Bornmann [10] and Bornmann [11] used data from this peer review system to investigate whether papers with the tag "good for teaching" achieve higher altmetrics counts than papers without this tag. Faculty members attach this tag to papers which indicates whether a paper could be of interest for other segments of society (i. e., the educational sector).

In the first step of this study, we analyze the underlying dimensions of measurement for traditional metrics (citation counts) and altmetrics – by using principal component analysis (PCA) and factor analysis (FA). In the second step, we test the relationship between the dimensions and the quality of papers (as measured by F1000Prime assessments) – using regression analysis. Only metrics should be used in research evaluation when they are related to the quality of papers.

## 2 Previous literature

Although the field of altmetrics has been established in scientometrics very recently, some literature reviews have been published by Bornmann [2], Thelwall [13], and Sugimoto et al. [14]. Many empirical studies have appeared which calculated the correlation between citations and altmetrics. The meta-analysis by Bornmann [15] summarizing the results of the studies show that "the correlation with traditional citations for micro-blogging counts is negligible (pooled $r = 0.003$), for blog counts it is small (pooled $r = 0.12$) and for bookmark counts from online reference managers, medium to large (CiteULike pooled $r = 0.23$; Mendeley pooled $r = 0.51$)" (p. 1123). Although many studies on altmetrics addressed the relationship of citations and altmetrics, we found only two studies focusing on the relationship of altmetrics and quality assessments by peers. In these studies, not quality scores from F1000Prime were used, but scores from the UK Research Excellence Framework (REF):

(1) The study by HEFCE [16] is based on data from REF 2014. The authors correlated different metrics with REF output quality profiles. "Quality profile" means a score of 0 (unclassified), 1, 2, 3, or 4 (best score) for individual publications in the correlation analysis. These scores have not been aggregated to the department level for the correlation analyses but are applied to individual publications. The correlations include a broad range of traditional and alternative metrics. The results are very different depending on the indicator and field. In the context of this study, we would like to focus on two tables from the report which focus on the correlation between REF quality scores and Mendeley readership as well as REF quality scores and Twitter counts. The overall correlation between REF quality scores and Mendeley readership is $r = 0.19$. The highest field-specific coefficients are reported for Clinical Medicine ($r = 0.44$) and Biological Sciences ($r = 0.36$), the lowest for Philosophy ($r = -0.07$) and Social Work and Social Policy ($r = -0.01$). In the case of Twitter counts, the overall correlation coefficient with REF quality scores is lower ($r = 0.07$) than that between Mendeley readership and REF quality scores ($r = 0.19$). The highest field-specific correlation coefficients can be found in Art and Design: History, Practice and Theory ($r = 0.23$) and Earth Systems and Environmental Sciences ($r = 0.21$), the lowest in Music, Drama, Dance and Performing Arts ($r = -0.07$) and Aeronautical, Mechanical, Chemical and Manufacturing Engineering ($r = -0.05$).

(2) The study by Ravenscroft et al. [17] focused on the references cited in case studies and correlated the altmetric scores for the references with the REF scores concerning societal impact. British universities use case studies in the REF to demonstrate societal impact. Ravenscroft et al. [17] used the Altmetric API to append the Altmetric attention score to the referenced publications in case studies. The Altmetric attention score is a weighted count including a broad range of different altmetrics (e.g., tweets, blog mentions). The authors found, however, for only around half of the publications the Altmetric attention score. Ravenscroft et al. [17] visualized the relationship between REF scores and Altmetric attention score and calculated the Pearson correlation coefficient. The very low and negative coefficient ($r = -0.0803$) points out that both scores seem to measure different constructs. It seems that the congruent validity of altmetrics with another indicator measuring societal impact is not given.

Whilst HEFCE [16] mirrors the approach of the current study, the study by Ravenscroft et al. [17] is a more indirect approach in this context: the study by Ravenscroft et al. [17] is based on cited publications in case studies submitted to the REF 2014.

# 3 Methods

## 3.1 Dataset used

We merged four different data sources:

(1) The papers selected for F1000Prime are rated by the Faculty members as "Good", "Very good", or "Exceptional" which is equivalent to scores of 1, 2, or 3, respectively. In many cases a paper is assessed not just by one member but by several [18]. The FFa (F1000 Article Factor), given as a total score in the F1000 database, is the sum of the scores from the different recommendations for a publication. For this study, we retrieved F1000Prime recommendation data ($n_r = 178,855$ recommendations for $n_p = 140,240$ papers, of which 131,456 papers have a digital object identifier, DOI) in November 2016.

(2) The CiteScore was downloaded from https://journalmetrics.scopus.com/ on January 12, 2017. CiteScore in year x counts the citations received in year x to documents published in years x-1, x-2, and x-3, and divides this by the number of documents published in years x-1, x-2, and x-3 (see https://journalmetrics.scopus.com).

(3) Citation counts from Web of Science (WoS) and Scopus were retrieved from our in-house databases developed and maintained by the Max Planck Digital Library (MPDL, Munich) and the Competence Center for Bibliometrics (http://www.bibliometrie.info/). The WoS citation counts were both derived from the Science Citation Index Expanded (SCI-E),

Social Sciences Citation Index (SSCI), Arts and Humanities Citation Index (AHCI) provided by Clarivate Analytics (formerly Thomson Reuters, Philadelphia, Pennsylvania, USA). Our in-house databases also contain Journal Impact Factor (JIF) values which were appended to the CiteScore values.

(4) Altmetrics data were used from a locally maintained database with data shared with us by the company Altmetric on June 04, 2016. The data include altmetrics from the following areas [see 19]: social networking (e.g., LinkedIn and Facebook counts), social bookmarking and reference management (e.g., Mendeley reader counts), social data sharing (e.g., Figshare), blogging (e.g., ScienceBlogs), microblogging (e.g., Twitter counts), wikis (e.g., Wikipedia), and social recommending (e.g., reddit). The data also include the Altmetric attention score (see https://help.altmetric.com/support/solutions/articles/6000060969-how-is-the-altmetric-attention-score-calculated) – a composite indicator – which is based on three factors: "(1) volume – the overall extent to which people are mentioning an article; (2) sources – types of places where the article is mentioned, some of which are valued more highly due to audience or prestige; and (3) authors – who is mentioning the article and to what extent these authors may be biased or engaged with scholarship" [20]. A critique of the score has been published by Gumpenberger et al. [21].

The altmetrics and F1000Prime data were matched with citation data via the DOI. Papers without a DOI were excluded from the analysis. Our download of the CiteScore data contained only data for publication years between 2011 and 2015. Citation counts were aggregated until the end of 2015. Therefore, only the publication years 2011-2013 were included to ensure a citation window of at least three years. Papers not found in the altmetrics database were also excluded. For 132 additional papers, no JIF was available in our in-house database. Thus, these papers were excluded, too. In total, 33,683 papers were included in the analysis.

Although the analyzed dataset is a significantly reduced dataset compared to the initial F1000Prime dataset (see above), we do not expect biased results. We do not see any relationship between the factors leading to the reduction and the results reported in this study. The main factor which reduced our dataset is the restriction to publication years between 2011 and 2013 which is, however, reasonable: (1) papers published after 2013 had too few time to accumulate citation counts and (2) usage of altmetrics like Twitter for academic purpose was less common before 2011.

From the initial altmetrics dataset with many different indicators, only a small set of indicators could be included in this study (namely, Twitter counts and the Altmetric attention score). Only these indicators have the necessary variance for inclusion in the PCA and FA. The other indicators are inflated by zero counts, which led to variances of zero or close to zero. Table 1 presents the number of papers and average values of the indicators included in this study per publication year.

Table 1. Average values of the indicators included in this study per publication year

| **Publication year**        | **2011** | **2012** | **2013** |
|-----------------------------|---------:|---------:|---------:|
| 3 years citations, WoS      | 29.69    | 29.93    | 29.95    |
| Citations, WoS              | 59.04    | 47.93    | 33.41    |
| 3 years citations, Scopus   | 33.91    | 33.38    | 25.77    |
| Citations, Scopus           | 58.62    | 42.51    | 26.24    |
| F1000 Score                 | 1.91     | 2.00     | 2.21     |
| CiteScore                   | 7.10     | 7.30     | 7.24     |
| Journal Impact Factor (JIF) | 10.67    | 11.20    | 11.43    |
| Twitter                     | 10.71    | 14.99    | 21.38    |
| Mendeley readers            | 84.57    | 75.82    | 68.97    |
| Altmetric attention score   | 9.54     | 15.00    | 26.83    |
| Number of papers            | 11,128   | 11,383   | 11,172   |

### 3.2 Statistical procedures

In this study, PCA is used to transform the different indicators considered here (e.g. bibliometric indicators) into new, uncorrelated variables [22]. The new variables are named as principal components whereby each component is a linear combination of the indicators. It is our objective to reduce the dimensionality of the indicators without losing as few as possible information. This can be achieved by selecting and interpreting only the first few components. The remaining components convey only a small amount of information. When there are a lot of indicators available, the use of the PCA is attractive, since the resulting components are not inter-correlated. Thus, the components can be used in a first step of analysis to understand the inter-relationships between the indicators. In a second step, the components can be included in a regression analysis to study the correlation between the components and quality assessments by peers. The use of the principal components overcomes the problem of multi-collinearity – the components are uncorrelated.

The study of Verardi [23] demonstrates that the results of a PCA – based on institutional performance data – can be distorted if the data are affected by outliers. Indeed, the underlying data of this study (bibliometric and altmetrics data) do not follow the normal distribution and are affected by outliers. In order to tackle this problem, on the one hand we logarithmized the indicator values by using the formula $\log_e(x+1)$. This logarithmic transformation has the effect that the data distributions approximate normal distributions. On the other hand, we performed the PCA using the robust covariance matrix following Verardi and McCathie [24]. Thus, the PCA in this study is not based on the indicator variables, but on a covariance matrix.

In addition to the PCA, we performed a FA to analyze the metrics. PCA and FA are similar statistical methods for data reduction [25]. In this study, the results from the FA are used to validate the results from the PCA – using a different method of data reduction. We calculated the FA by using the robust covariance matrix (see above) which has been transformed into a correlation matrix [26]. We used the principal-component factor method to analyze the correlation matrix; the communalities are assumed to be 1. We interpreted the factor loadings for the orthogonal varimax rotation; the factor loadings have been adjusted

"by dividing each of them by the communality of the correspondence variable. This adjustment is known as the Kaiser normalization" [22]. In the interpretation of the results, we focused on factor loadings with values greater than 0.5.

Besides analyzing the underlying dimensions of the indicators, we were interested in the correlation between the dimensions and F1000Prime scores (as proxies for the quality of papers). The scores are a count variable which is a weighted number of recommendations from F1000Prime faculty members. Thus, the variable is the sum of the F1000Prime recommendations for single papers. For example, if a paper receives the recommendations 3, 2, and 1, the summarized F1000Prime score is 6. For the calculation of the correlations between the results of PCA (and FA) and F1000Prime scores, we performed a robust, negative binomial regression (NBREG) [27,28]. The NBREG model is recommended to be used in case of over-dispersed count data (where the variance exceeds the mean) [29]. Robust methods are recommended when the distributional assumptions for the model are not met [28], as it is the case for citation and altmetrics data.

Similar to our approach in data reduction (see above), we calculated further (robust) regression analyses to validate the results of the NBREG. We additionally performed ordinary least squares (OLS) regression analyses based on logarithmized indicator values – by using the formula $\log_e(x+1)$ – following the recommendations by Thelwall and Wilson [30].

# 4 Results

## 4.1 Principal components analysis (PCA) and factor analysis (FA)

The reduction of dimensionality is one important objective of PCA. Table 2 shows the components as the results of the PCA using the metrics data. The components in the table are arranged in decreasing order of variance. Thus, the first component is able to explain 72% of total variance in the metrics. The cumulative proportion of total variance indicates how much information is retained by choosing a certain number of components. Various rules exist how many components should be selected for further analyses [31]. According to Afifi et al. [22], there is one common cut-off point at a proportion of 80%. This level is reached with two components in Table 2.

Table 2. Eigenvalues and (cumulative) proportions of total variance for metrics data (n=33,683)

| Component | Eigenvalue | Proportion | Cumulative proportion |
|---|---|---|---|
| **Comp1** | **6.28** | **72%** | **72%** |
| **Comp2** | **1.13** | **13%** | **84%** |
| Comp3 | 0.71 | 8% | 93% |
| Comp4 | 0.46 | 5% | 98% |
| Comp5 | 0.10 | 1% | 99% |
| Comp6 | 0.05 | 1% | 99% |
| Comp7 | 0.02 | 0% | 100% |
| Comp8 | 0.02 | 0% | 100% |
| Comp9 | 0.01 | 0% | 100% |

Table 3 shows the correlation coefficients for the first two principal components and the metrics data. Following the guidelines by Afifi et al. [22], the indicators with a correlation greater than 0.5 are emphasized. Thus, the correlation is formatted in bold font face if it exceeds 0.5/sqrt(eigenvalue). We formatted two correlations in bold face and in italics, because they are marginally below the cut-off point. Since we can expect that there are measurement errors in the bibliometric data which add stochastic elements to the analysis [32], we interpret these correlations as meaningful too.

Table 3. Principal components analysis for metrics data (n=33,683)

| **Variable (logarithmized)** | **Component 1** | **Component 2** |
|---|---|---|
| 3 years citations, WoS | **0.41** | -0.08 |
| Citations, WoS | **0.46** | -0.18 |
| Citations, Scopus | **0.46** | -0.25 |
| 3 years citations, Scopus | **0.42** | -0.15 |
| Tweets | 0.09 | **0.73** |
| Mendeley readers | **0.40** | 0.31 |
| Altmetric attention score | 0.05 | 0.35 |
| CiteScore | ***0.17*** | 0.23 |
| Journal Impact Factor (JIF) | **0.20** | 0.25 |
| | | |
| Eigenvalues | 6.28 | 1.13 |
| Cumulative proportion | 0.72 | 0.84 |
| **Cut-off point: 0.5/sqrt(eigenvalue)** | **0.20** | **0.47** |

As the results in the table indicate, citations for single papers and journals as well as Mendeley readers are highly correlated with the first component. A high value of the first component is an indication that a paper was published in a high impact journal and received a lot of reader and citation impact. Instead, the second component in Table 2 reflects Twitter counts – one of the most frequently used sources for altmetrics studies in scientometrics. The Altmetric attention score is not or only scarcely related to the second component. This result might reflect that it is a composite score including a broad range of altmetrics (although the Altmetric attention score is mainly driven by Twitter counts).

The results of the FA are shown in Table 4 which validate the main result of the PCA. The Kaiser criterion suggested to retain three factors in the analysis with eigenvalues equal to or higher than 1. Although this is one more dimension than in the PCA, the additional dimension is determined by journal metrics which were in the PCA at the threshold of consideration in the first dimension. The second difference between the results of the PCA and FA concerns the Altmetric attention score. The score loads on the same factor as Twitter counts, which is not the case in the PCA. However, the loadings of the Altmetric attention score in the PCA are rather close to, but significantly below the recommended threshold.

Table 4. Factor analysis for metrics data (n=33,683)

| **Variable (logarithmized)** | **Factor 1** | **Factor 2** | **Factor 3** |
|---|---|---|---|
| 3 years citations, WoS | **0.93** | 0.27 | 0.12 |
| Citations, WoS | **0.95** | 0.26 | 0.04 |
| Citations, Scopus | **0.96** | 0.21 | 0.00 |
| 3 years citations, Scopus | **0.96** | 0.22 | 0.08 |
| Tweets | 0.07 | 0.11 | **0.98** |
| Mendeley readers | **0.62** | 0.48 | 0.17 |
| Altmetric attention score | 0.08 | 0.10 | **0.98** |
| CiteScore | 0.28 | **0.94** | 0.11 |
| Journal Impact Factor (JIF) | 0.31 | **0.92** | 0.11 |
| | 0.93 | 0.27 | 0.12 |
| Variance | 4.18 | 2.21 | 2.00 |
| Cumulative proportion | 0.46 | 0.71 | 0.93 |

Added together, the metrics seem to reflect two (or three) dimensions: The first one reflects impact on academia (readers and citers). The results of the PCA reveal that the dimension is partly dependent on the citation impact of the publishing journals. Similar results from another factor analysis have been published by Zahedi et al. [9]: "citation indicators are more correlated between them than with altmetrics" (p. 1505). However, a common dimension for journal metrics and metrics on the level of single papers does not accord to the results of our factor analysis and the factor analysis by Costas et al. [33], which both found two independent dimensions. The second component (PCA) and third factor (FA) in our study seems to reflect the wider impact (beyond science) which is largely independent from academic impact. It is mainly based on Twitter counts.

## 4.2   Negative binomial regression analysis (NBREG)

In the second step of the statistical analysis in this study, we were interested in the correlation between the quality of papers (measured by FFa) and the two or three impact dimensions from the PCA or FA, respectively. Thus, we included the FFa as dependent and the scores for the impact dimensions as independent variables in the NBREG. We calculated two NBREG, which included either the components from the PCA or the factors (unrotated results) from the FA. Both results are shown in Table 5. The coefficients of all independent variables are statistically significant; however, the number of papers is very high [34].

For the interpretation of the correlation between FFa and the dimensions, the marginal effects – as a measure of practical significance – are more interesting [27,35]. The effects for the components from the PCA in Table 5 can be interpreted as follows: An average increase of a paper's expected FFa by 0.62 is related to a standard deviation change in the first component (citation and reader impact). However, the same change in the second component (mainly Twitter counts) is only related to an average increase of a paper's expected FFa by 0.3. In other words, the first component seems to be significantly stronger related to the quality of papers in practical terms than the second component.

The results of the NBREG in Table 5 which are based on the factor scores from the FA validate the results based on the PCA components. The first factor (citation and reader impact) is strongly related to the quality of papers, but the third factor (tweets and the Altmetric attention score) is not. The difference between both factors is still larger than that in the NBREG based on PCA components. We assume that the difference is larger in the FA than in the PCA because an additional factor reflecting journal impact exists (factor 2) which is similarly scarcely related to the quality of papers as the altmetrics factor.

Table 5. Beta coefficients of and marginal effects from two negative binomial regression analyses (NBREG, n=33,683)

|  | F1000Prime score | Marginal effects (+SD) |
|---|---|---|
| Scores from principal components analysis |  |  |
| Scores for component 1 | 0.11*** | 0.62 |
|  | (51.83) |  |
| Scores for component 2 | 0.09*** | 0.30 |
|  | (29.67) |  |
| Constant | -0.31*** |  |
|  | (-20.72) |  |
| Scores from factor analysis |  |  |
| Scores for factor 1 | 0.30*** | 0.75 |
|  | (55.94) |  |
| Scores for factor 2 | 0.06*** | 0.23 |
|  | (19.62) |  |
| Scores for factor 3 | 0.09*** | 0.19 |
|  | (18.04) |  |
| Constant | -0.46*** |  |
|  | (-28.66) |  |

Notes. *t* statistics in parentheses

*** $p < 0.001$

We additionally calculated two OLS regression analyses to validate the results of the NBREGs. The results from two analyses, one including the scores from PCA and the other including the scores from FA are presented in Table 6. Although the difference between citation impact (on the single paper level) and altmetrics is less pronounced than in Table 5, the different relationships between both metrics and the quality of papers are similarly visible as in Table 5.

Table 6. Beta coefficients of and marginal effects from two ordinary least squares regression analyses (OLS regression, n=33,683)

|  | F1000Prime score | Marginal effects (+SD) |
|---|---|---|
| Scores from principal components analysis |  |  |
| Scores for component 1 | 0.05*** | 0.13 |
|  | (54.07) |  |
| Scores for component 2 | 0.06*** | 0.09 |
|  | (36.81) |  |
| Constant | 0.52*** |  |
|  | (77.19) |  |
| Scores from factor analysis |  |  |
| Scores for factor 1 | 0.64*** | 0.66 |
|  | (46.10) |  |
| Scores for factor 2 | 0.15*** | 0.29 |
|  | (21.31) |  |
| Scores for factor 3 | 0.17*** | 0.17 |
|  | (15.52) |  |
| Constant | -0.35*** |  |
|  | (-8.85) |  |

Notes. $t$ statistics in parentheses
*** $p < 0.001$

## 5    Discussion

According to the NISO Alternative Assessment Metrics Project [36] altmetrics can possibly play a role in research evaluation in the assessment of non-academic impact: how is the evaluated unit engaged with the social, cultural, and economic environment. This engagement can be seen as important; it represents 20% of the evaluation in the current version of the REF in the UK. There is also another aspect where altmetrics can play a role: the assessment of research outputs which have not traditionally been part of evaluations, e.g., research data or scientific software. It is the intention of the current study to test whether altmetrics can really be used in research evaluation practice. If metrics are not related to scientific quality, there is the risk that bad or fraudulent research is rated high, because it simply received attention in society [37].

We calculated both PCA and FA to reduce the metrics variables to few dimensions (components or factors). The results of the PCA show that citations and reads operate along similar dimensions, but tweets along different dimensions. This result is in agreement with the meta-analysis of Bornmann [15] and similar to the results of the PCA of Zahedi et al. [9]. Whereas the factor analysis of Costas et al. [33] separates between altmetrics and citation impact metrics (they did not consider Mendeley counts), the journal-based metrics load on a different factor than the paper-based citation impact metrics. This result of Costas et al. [33] accords to the results which we received from our FA.

We included scores from the PCA and FA in two NBREGs. The results of both robust NBREG in this study indicate that citations and reads are significantly stronger related to quality (as measured by FFa) than tweets (and the Altmetric attention score). The small correlation with tweets accords to the results of HEFCE [16]. The reported Spearman coefficient for the correlation of Twitter counts and REF quality scores amounted to $r_s=0.07$.

These results might question the use of tweets (and the Altmetric attention score) for research evaluation purposes but indicates potential use of Mendeley reader counts. However, one should have in mind that citations and reads stronger reflect impact by (prospective) researchers than tweets. Tweets are frequently written by people working outside the science sector. The analysis of Twitter user categories by Yu [38] shows that "the general tweet distribution is strongly influenced by the public user" (p. 267). Since one cannot expect that non-experts assess the quality of papers properly, research evaluations using Twitter data should only include papers fulfilling certain quality standards. The same should (probably) be done, if the Altmetric attention score is included in an evaluation.

Such a procedure is, for example, considered in the broad impact measurement of research in the REF: "The REF process advised (through guidelines and people associated with the process) that all types of impact beyond academia were admissible, but that the impact had to benefit a particular sector and include beneficiaries outside of the academe as well as link directly back to published academic work of reasonably high international standard ('two-star' minimum)" [39]. "Two-star" means "quality that is recognised internationally in originality, significance and rigour" (see www.ref.ac.uk).

We do not want to argue with these proposals against measurements of public opinion or attention of papers in scientometrics. These investigations allow interesting insights in the popularity of research topics and themes. However, we would like to point with our study to the general necessity of considering scientific quality standards in research evaluations (using altmetrics).

This study is based on a dataset with papers recommended by F1000Prime. The dataset is restricted to the biomedical area. Furthermore, the results of Bornmann [18] show that F1000Prime mostly covers publications with high quality level – measured in terms of citation impact. Since the results cannot be generalized to other disciplines, such as physics or social sciences, and papers with lower quality levels, we encourage future studies including assessments by peers, focusing on other disciplines, and using more diverse datasets according to the quality of papers.

# Acknowledgements

This paper is based on a contribution, which has been presented at the Science, Technology, and Innovation Indicators conference (STI 2017) in Paris, France.

The bibliometric data (WoS and Scopus) used in this paper are from in-house databases developed and maintained by the Max Planck Digital Library (MPDL, Munich) and the Competence Center for Bibliometrics (http://www.bibliometrie.info). The WoS data are derived from the Science Citation Index Expanded (SCI-E), Social Sciences Citation Index (SSCI), and Arts and Humanities Citation Index (AHCI) provided by Clarivate Analytics (http://clarivate.com). The altmetrics data were taken from a data set retrieved from Altmetric on June 04, 2016 and stored in a local database and maintained by the Max Planck Institute for Solid State Research (Stuttgart). The F1000Prime data were received from F1000 in November 2016. The complete dataset excluding DOIs have been made available at https://doi.org/10.6084/m9.figshare.6120158.v1